\documentclass[twocolumn,showpacs,preprintnumbers,amsmath,amssymb,superscriptaddress]{revtex4}

\usepackage{graphicx}
\begin{document}
\bibliographystyle{prsty}
\title{Unveiling the impurity band inducing ferromagnetism in magnetic semiconductor (Ga,Mn)As
}

\author{Masaki~Kobayashi}
\altaffiliation{Present address: Photon Factory, 
Institute of Materials Structure Science, High Energy Accelarator Research Organization, 
Tsukuba, Ibaraki 305-0801, Japan}
\affiliation{Department of Applied Chemistry, 
School of Engineering, University of Tokyo, 
7-3-1 Hongo, Bunkyo-ku, Tokyo 113-8656, Japan}
\affiliation{Swiss Light Source, Paul Scherrer Institut, 
CH-5232 Villigen PSI, Switzerland}
\author{Iriya~Muneta}
\affiliation{Department of Electrical Engineering and Information Systems, 
University of Tokyo, 7-3-1 Hongo, Bunkyo-ku, Tokyo 113-8656, Japan}
\author{Yukiharu~Takeda}
\affiliation{Synchrotron Radiation Research Unit, 
Japan Atomic Energy Agency, Sayo-gun, Hyogo 679-5148, Japan}
\author{Yoshihisa~Harada}
\affiliation{Institute for Solid State Physics, 
University of Tokyo, Kashiwanoha, Kashiwa, Chiba 277-8561, Japan}
\author{Atsushi~Fujimori}
\affiliation{Department of Physics, University of Tokyo, 
7-3-1 Hongo, Bunkyo-ku, Tokyo 113-0033, Japan}
\author{Juraj~Krempask{\'{y}}}
\affiliation{Swiss Light Source, Paul Scherrer Institut, 
CH-5232 Villigen PSI, Switzerland}
\author{Thorsten~Schmitt}
\affiliation{Swiss Light Source, Paul Scherrer Institut, 
CH-5232 Villigen PSI, Switzerland}
\author{Sinobu~Ohya}
\affiliation{Department of Electrical Engineering and Information Systems, 
University of Tokyo, 7-3-1 Hongo, Bunkyo-ku, Tokyo 113-8656, Japan}
\author{Masaaki~Tanaka}
\affiliation{Department of Electrical Engineering and Information Systems, 
University of Tokyo, 7-3-1 Hongo, Bunkyo-ku, Tokyo 113-8656, Japan}
\author{Masaharu~Oshima}
\affiliation{Department of Applied Chemistry, 
School of Engineering, University of Tokyo, 
7-3-1 Hongo, Bunkyo-ku, Tokyo 113-8656, Japan}
\author{Vladimir~N.~Strocov}
\affiliation{Swiss Light Source, Paul Scherrer Institut, 
CH-5232 Villigen PSI, Switzerland}
\date{\today}

\begin{abstract}
(Ga,Mn)As is a paradigm diluted magnetic semiconductor which shows ferromagnetism induced by doped hole carriers. With a few controversial models emerged from numerous experimental and theoretical studies, the mechanism of the ferromagnetism in (Ga,Mn)As still remains a puzzling enigma. 
In this Letter, we use soft x-ray angle-resolved photoemission spectroscopy to positively identify the ferromagnetic Mn $3d$-derived impurity band in (Ga,Mn)As. The band appears hybridized with the light-hole band of the host GaAs. 
These findings conclude the picture of the valence band structure of (Ga,Mn)As disputed for more than a decade. The non-dispersive character of the IB and its location in vicinity of the valence-band maximum indicate that the Mn $3d$-derived impurity band is formed as a split-off Mn-impurity state predicted by the Anderson impurity model. 
Responsible for the ferromagnetism in (Ga,Mn)As is the transport of hole carriers in the impurity band. 
\end{abstract}


\maketitle

Ferromagnetic diluted magnetic semiconductors (DMS) are formed by substitution of several percent of cation sites in a host semiconductor by magnetic impurities. Because the carriers in DMS are considered to mediate the magnetic interaction between the magnetic ions \cite{NatMater_10_Dietl}, these materials are of key importance for ``spintronics'' aiming at development of advanced functional devices to control the spin degree of freedom of the carriers. 
Due to the mediation mechanism, the ferromagnetism in DMS is called carrier-induced ferromagnetism. This mechanism arms with the capability of manipulating both the electron charge and spin degrees of freedom in functional spintronic devices. 
The III-V based DMS (Ga,Mn)As is a prototype ferromagnetic DMS, which has been intensively studied from both the fundamental and applicational points of view \cite{RMP_06_Jungwirth}. 
Although the Curie temperature ($T_\mathrm{C}$) of (Ga,Mn)As is at present lower than the room temperature, applications of (Ga,Mn)As to functional spintronic devices have been established.

Understanding the mechanism of the carrier-induced ferromagnetism is of paramount importance for further development of the spintronic device applications with DMS. Several theoretical models for the carrier-induced ferromagnetism in (Ga,Mn)As have been proposed \cite{NatMater_10_Dietl}. 
In the limit where the hole carriers are considered nearly free or itinerant, the Zener $p$-$d$ exchange model has been suggested based on mean-field theory point of view \cite{Science_00_Dietl, PRB_07_Jungwirth}. 
Here, the Mn $3d$ acceptor level is located above the Fermi level ($E_\mathrm{F}$) and merges into the valence band (VB) of the host GaAs through its hybridization, leading to the exchange-split VB of host GaAs crossing $E_\mathrm{F}$. This model can explain the carrier-concentration dependence of $T_\mathrm{C}$ measured on an electric-field transistor structure using (Ga,Mn)As \cite{NatPhys_10_Sawicki}. In the opposite limit where the hole carriers are strongly localized around the magnetic impurity, the Mn $3d$ impurity band (IB) model has been proposed \cite{PRL_01_Berciu, APL_04_Mahadevan}. 
Here, in contrast to the Zener model, the IB intersects $E_\mathrm{F}$. This model can explain the experimental results of vacuum-ultraviolet angle-resolved photoemission spectroscopy (ARPES) \cite{PRB_01_Okabayashi}, hard x-ray ARPES \cite{NatMater_12_Gray} and optical spectroscopy suggesting that $E_\mathrm{F}$ resides in the IB \cite{PRL_06_Burch}, as well as the results of resonant tunneling spectroscopy indicating that $E_\mathrm{F}$ is located above valence-band maximum (VBM) of the host GaAs and the GaAs valence states remain mostly unchanged \cite{NatPhys_11_Ohya, PRL_10_Ohya}. 
Additionally, as a model bridging from the localized hole carriers to the delocalized hole carriers causing the ferromagnetism, percolation theory of bound magnetic polarons (BMPs) has been proposed \cite{PRL_02_Kaminski, PRB_03_Kaminski}. The model well explains the temperature dependence of the transport property of (Ga,Mn)As with low-carrier concentration \cite{PRB_03_Kaminski}, and is likely relevant to an understanding of the nanoscale phase separation \cite{NatPhys_10_Sawicki} and inhomogeneous growth of magnetic domains near $T_\mathrm{C}$ \cite{PRL_08_Takeda}.

A Ga$_{1-x}$Mn$_x$As ($x = 0.025$) thin film with a thickness of 100 nm was grown on a GaAs(001) substrate at 275 $^{\circ}$C under ultra-high vacuum by molecular beam epitaxy method. 
To avoid surface oxidation, after the deposition of Ga$_{1-x}$Mn$_x$As layer the sample was covered by an amorphous As capping layer, to produce a structure As/Ga$_{0.975}$Mn$_{0.025}$As/GaAs(buffer)/GaAs(001). 
The Curie temperature $T_{\mathrm{C}}$ of our sample was $\sim 35$ K as determined by the Arrott plot of the magnetic circular dichroism. The SX-ARPES experiments were conducted at the SX-ARPES end station of the ADRESS beamline in Swiss Light Source \cite{JSR_10_Strocov}. 
In the $h\nu$ range from 350 to 1000 eV, the combined beamline and analyzer energy resolution including the thermal broadening varied from 50 to 150 meV. The analyzer slit was oriented to lie in the measurement plane. The measurements used linear vertical ($p$) and horizontal ($s$) polarizations of incident beam. The measurements were performed at a temperature of 11 K under ultrahigh vacuum of $5.0 \times 10^{-11}$ mbar. The Mn $L_{2,3}$ XAS spectra were measured in the total-electron-yield mode.

Our Ga$_{1-x}$Mn$_x$As ($x = 0.025$) thin-film sample was capped by an amorphous As layer, which well protects a GaAs underlayer against oxidization, exposure to deionized water, and annealing up to 180 $^{\circ}$C \cite{EL_84_Kawai}. 
In this report, we demonstrate that the use of soft x-ray (SX) ARPES with photon energies $h\nu$ towards 1 keV allows the observation of the bulk band dispersion of the Ga$_{1-x}$Mn$_x$As underlayer through the amorphous passivation layer, with any contributions from their interface being hardly seen in the spectra \cite{APL_12_Kobayashi}. Hereafter, we shall focus on the bulk electronic structure of (Ga,Mn)As seen in our SX-ARPES experiment.

\begin{figure}[t!]
\begin{center}
\includegraphics[width=8.8cm]{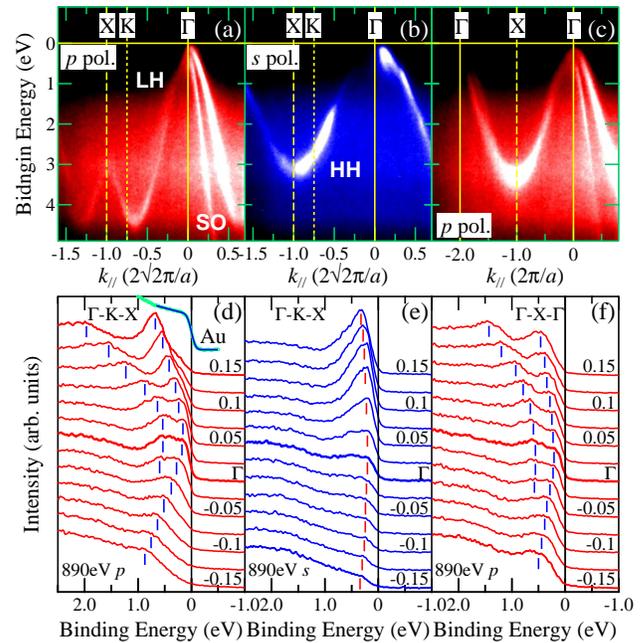}
\caption{Band dispersion near the $\Gamma$ point in (Ga,Mn)As. 
The incident photon energy is $h\nu = 890$ eV. 
(a),(b) ARPES images along the $\Gamma$-K-X line measured with $p$- and $s$-polarizations, respectively. 
The light-hole (LH) and split-off (SO) bands are clearly seen with the $p$ polarization, while the heavy-hole (HH) band is active with the s polarization. 
(c) ARPES image along the $\Gamma$-X-$\Gamma$ line. 
(d)-(f) Energy-distribution curves (EDCs) around the $\Gamma$ point corresponding to (a)-(c). 
The vertical bars are guides for the eyes tracing the spectra structures. 
The Fermi level position has been measured with Au foil in electrical contact with the sample.
}
\label{GMA_BD}
\end{center}
\end{figure}

\begin{figure*}[t!]
\begin{center}
\includegraphics[width=18cm]{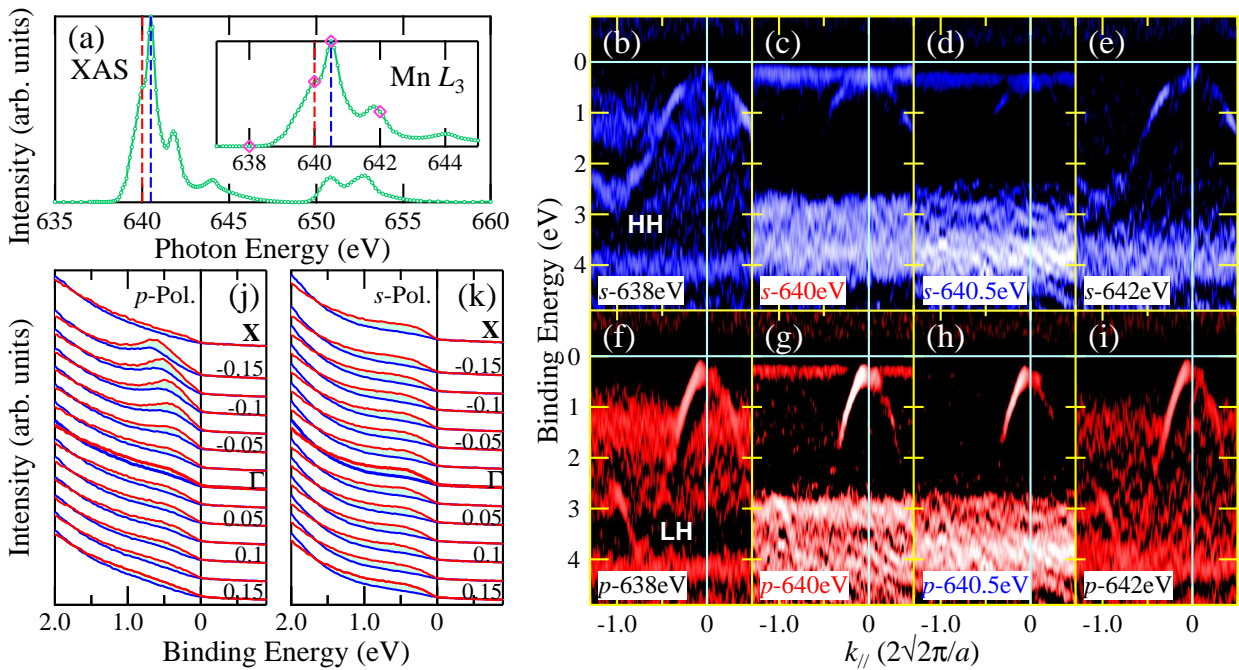}
\caption{Resonant angle-resolved photoemission spectroscopy (r-ARPES) spectra at the Mn $L_{2,3}$ edge in (Ga,Ma)As. 
(a) Mn $L_{2,3}$ XAS spectrum. The red and blue vertical dashed lines correspond to the resonant $h\nu$ for the ferromagnetic and paramagnetic components, respectively. 
(b)-(e), (f)-(i) r-ARPES images taken in a series of at $h\nu$ from 638 to 642 eV with $s$- and $p$-polarization, respectively, with the $h\nu$ points are shown as rhombi in the inset of Fig.(a). 
All the images are represented in 2nd derivative along $E_B$ and the intensities are plotted in logarithmic scale to emphasize the resonance enhancements. 
(j),(k) Resonant EDCs around the $\Gamma$ and at the X points taken with $p$- and $s$-polarizations, respectively. 
The red curves are the EDCs for $h\nu = 640$ eV, the blue ones for $h\nu = 640.5$ eV. The green areas denote differences between the red and blue curves.
}
\label{GMA_AR2PES}
\end{center}
\end{figure*}

Figure~\ref{GMA_BD} shows the binding energy ($E_B$) {\it vs}. momentum ($k$) plots along the $\Gamma$-K-X symmetry line of the Brillouin zone measured with $p$- and $s$-polarizations at $h\nu$ of 890 eV. Due to the wavefunction symmetry properties, the light-hole (LH) and split-off (SO) bands of the host GaAs show up with the $p$-polarization [Fig.~\ref{GMA_BD}(a)], whereas the heavy-hole (HH) band pops up in the image taken with the $s$-polarization [Fig.~\ref{GMA_BD}(b)]. 
The LH and HH bands become almost degenerated along the $\Gamma$-X-$\Gamma$ line, as shown in Fig.~\ref{GMA_BD}(c). 
These dispersive bands form the band manifold characteristic of the GaAs band structure. Our results are qualitatively consistent with the previous ARPES experiments and band calculations on GaAs \cite{PRB_80_Chiang, PRL_09_Luo, NatMater_11_Gray}. 
Figures~\ref{GMA_BD}(d) and ~\ref{GMA_BD}(e) display the energy distribution curves (EDCs) along the $\Gamma$-K-X symmetry line around the $\Gamma$ point with $p$- and $s$-polarizations, respectively. 
Note that although the LH and HH bands approach $E_\mathrm{F}$ in the vicinity of the $\Gamma$ point, they do not intersect $E_\mathrm{F}$ even at their top. The same is observed along the $\Gamma$-X-$\Gamma$ line as shown in Fig.~\ref{GMA_BD}(c). 
These results set unambiguous evidence that the VBM of host GaAs is located below $E_\mathrm{F}$, which is consistent with the results of resonant tunneling spectroscopy \cite{NatPhys_11_Ohya} and immediately dismisses the Zener $p$-$d$ exchange model \cite{Science_00_Dietl, PRB_07_Jungwirth}. 
Furthermore, we note a small but finite spectral weight at $E_\mathrm{F}$, which is a direct observation of finite density of states at $E_\mathrm{F}$ in (Ga,Mn)As. 
This fact has been missed by all previous photoemission experiments including $in$-$situ$ ones \cite{PRB_01_Okabayashi, NatMater_12_Gray, PRB_02_Asklund, PRB_04_Rader} and is consistent with the metallic conductivity of (Ga,Mn)As. 
This observation warrants that the SX-ARPES measurements with their enhanced probing depth reflect the bulk electronic structure of (Ga,Mn)As. No signal from the Mn $3d$ states can be distinguished from the VB of host GaAs in these spectra taken at high $h\nu$.

The energy position of the Mn $3d$ states in VB has intensively been debated for more than a decade. To address this issue, we have conducted resonant angle-resolved photoemission spectroscopy (r-ARPES) at the Mn $L_3$ absorption edge. r-ARPES is known to probe the element-specific band dispersion of open-shell $d$ or $f$ electron systems \cite{PRL_08_Im, PRB_10_Mulazzi}. 
Figure~\ref{GMA_AR2PES}(a) shows the x-ray absorption spectroscopy (XAS) spectrum at the Mn $L_{2,3}$ edges of the sample. Previously, the magnetic-field ($H$) dependence of x-ray magnetic circular dichroism (XMCD) has demonstrated that (Ga,Mn)As includes two kinds of Mn components: The intrinsic one, which is most likely due to ferromagnetic substitutional Mn ions including part of paramagnetic interstitial Mn$^{2+}$ ions, and the extrinsic one, which is due to paramagnetic oxidized Mn$^{2+}$ ions segregated in the surface region \cite{PRL_08_Takeda, PRL_04_Edmonds}. 
The XAS peak position of the ferromagnetic intrinsic Mn component of 640 eV is different from that of the paramagnetic extrinsic one of 640.5 eV [red and blue vertical lines in Fig.~\ref{GMA_AR2PES}(a)] \cite{PRL_08_Takeda}. 
Accordingly, the resonance enhancement of ARPES measured at $h\nu = 640$ eV should be relevant to the intrinsic components, while the r-ARPES spectrum taken at $h\nu = 640.5$ eV should have almost zero contribution of the ferromagnetic states. Therefore, comparison between the r-ARPES spectra taken at 640 eV and 640.5 eV will reveal the Mn $3d$ states of the intrinsic origin which are responsible for the ferromagnetism.

Figures~\ref{GMA_AR2PES}(b)-\ref{GMA_AR2PES}(e) show a series of r-ARPES images measured on our (Ga,Mn)As sample under variation of $h\nu$ with $s$-polarization (no shift of the band dispersion could be seen in the spectra taken at different Brillouin zones). The resonance at $h\nu = 640$ eV also enhances the intensity around $E_B \sim 3$ eV, which is due to the well-known main peak of the Mn $3d$ partial density of states \cite{PRB_04_Rader}. 
Most important to note in the evolution of the r-ARPES images is the Mn $3d$-derived IB which pops up at $h\nu = 640$ eV in the vicinity of $E_\mathrm{F}$, as shown in Fig.~\ref{GMA_AR2PES}(c), due to the resonance enhancement at the excitation energy of the ferromagnetic intrinsic Mn component, not paramagnetic extrinsic Mn oxides. 
Conversely, the r-ARPES image in Fig.~\ref{GMA_AR2PES}(d) taken at $h\nu = 640.5$ eV, which resonates with the paramagnetic Mn component and is non-resonance for the intrinsic one, shows only an afterglow of the IB near $E_\mathrm{F}$ and another non-dispersive resonant projection occurs around $E_B \sim 4$ eV.
We note that the existence of the IB has been reported in previous ARPES measurements on (Ga,Mn)As with vacuum-ultraviolet rays ($h\nu = 18 -40$ eV) \cite{PRB_01_Okabayashi} and hard x-rays \cite{NatMater_12_Gray} but the location of the IB was deeper ($\sim 0.4$ eV below $E_\mathrm{F}$) than in our observation and overlapped with the VBM. 
Furthermore, theoretical calculations subsequent to the vacuum-ultraviolet ARPES \cite{PRB_01_Okabayashi} suggested that the IB originated from the interstitial Mn impurities \cite{PRL_05_Ernst}. The present observations give unambiguous evidence that the IB is in fact located near the VBM and formed by the intrinsic Mn ions directly related to the ferromagnetism in (Ga,Mn)As. Even though the intrinsic Mn components include not only the Mn ions substituting for the Ga site but also partly interstitial Mn ions antiferromagnetically coupled to the substitutional Mn, influence of the interstitial Mn is minor in the r-ARPES spectra because the amount of the interstitial Mn ions is only $\sim 10$\% of the total amount of Mn ions in (Ga,Mn)As \cite{NatMater_12_Dobrowolska}. 
We note that although the spectral weight seen at $E_\mathrm{F}$ is small, it must be non-zero because of its disordered character of the IB which introduces energy broadening of this band with its high-energy tail approaching $E_\mathrm{F}$.

In addition, incident light polarization dependence of the r-ARPES signal sheds light on details of the hybridization between the Mn $3d$ orbital and the ligand bands. 
Figure~\ref{GMA_AR2PES}(f)-\ref{GMA_AR2PES}(i) shows the Mn $L_3$ r-ARPES series of the (Ga,Mn)As sample taken with $p$-polarization. The off-resonance image taken at $h\nu = 638$ eV shows the LH band of host GaAs over the background of non-dispersive amorphous As states, as shown in Fig.~\ref{GMA_AR2PES}(f). The $h\nu = 640$ eV resonance of the ferromagnetic intrinsic Mn component enhances the Mn $3d$-derived IB in the vicinity of the VB maximum and the Mn $3d$ partial density-of-states around binding energy $E_B \sim 3$ eV, as shown in Fig.~\ref{GMA_AR2PES}(g). It is important to note here that together with the IB the $h\nu = 640$ eV resonance also pops up the LH band. 
Figures~\ref{GMA_AR2PES}(j) and \ref{GMA_AR2PES}(k) show the linear-polarization dependence of the EDCs around the $\Gamma$ and at the X points measured at the Mn $L_3$ resonance. With $p$-polarization, which exposes the LH band of host GaAs, the difference of the EDCs taken at 640 eV and 640.5 eV [the shaded area in Fig.~\ref{GMA_AR2PES}(j)] representing the resonant contribution disperses as a function of $k$, closely following the LH band. These manifest the IB hybridization with the LH band. In contrast, the EDCs taken with $s$-polarization, which exposes the HH band, show only the $k$-independent intensity of the IB, suggesting only an insignificant hybridization with the HH band. The results indicate that the IB hybridizes with the LH band but not with the HH one having different wavefunction character \cite{PRL_10_Ohya}.

\begin{figure}[t!]
\begin{center}
\includegraphics[width=8.8cm]{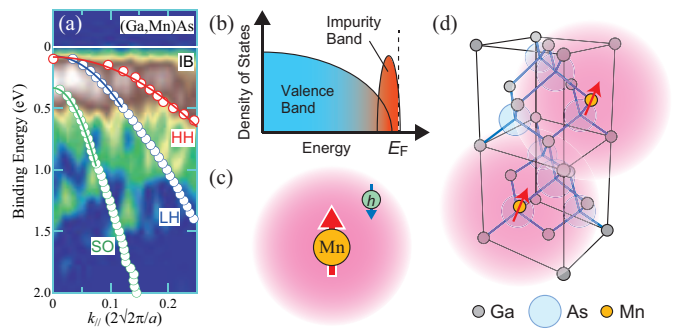}
\caption{Valence-band structure of (Ga,Mn)As. 
(a) Band dispersion around the $\Gamma$ point near $E_\mathrm{F}$ along the $\Gamma$-K-X line. IB denotes the Mn $3d$-derived impurity band. The circles are peak positions of the momentum-distribution curves (MDCs) fitted with the Lorenzian function, reflecting the HH, LH and SO bands. 
The solid lines are their fit with quadratic polynomial function centered at the $\Gamma$ point to the band dispersion. 
(b) Density of states of the Mn $3d$-derived IB embedded into the VB in host GaAs. 
(c) Zhang-Rice single-like bound magnetic polaron around the doped Mn ion. 
(d) Schematic image of overlap among the magnetic polarons in (Ga,Mn)As.
}
\label{GMA_VB}
\end{center}
\end{figure}

Figure~\ref{GMA_VB}(a) shows a summary of the VB structure of (Ga,Mn)As near the $\Gamma$ point. Here, the positions of the LH, SO, and HH bands were estimated from the Lorentzian fits of momentum-distribution curves (MDCs) of the experimental ARPES images. 
The obtained dispersive bands were fitted in turn with quadratic polynomial functions. 
The estimated positions of the LH, HH, and SO band maxima are at $E_B \sim 75$ meV, 85 meV, and 336 meV, respectively, providing us with further evidence that VB of the host GaAs is located below $E_\mathrm{F}$. 
Figure~\ref{GMA_VB}(b) shows a schematic diagram of the VB electronic structure of (Ga,Mn)As which embeds our main experimental findings: 
{\bf I}. Formation of a non-dispersive Mn $3d$-derived IB is mostly related to the substitutional Mn impurities. The absence of its dispersion demonstrates the random distribution of the Mn ions in (Ga,Mn)As, with the crystal momentum {\bf k} being not a good quantum number for this band. Therefore, the formation of the IB responsible for the ferromagnetism can be theoretically understood in the framework of local models. The central position of the IB is located below $E_\mathrm{F}$; 
{\bf II}. The VBM of host GaAs is also located below $E_\mathrm{F}$; 
{\bf III}. The Mn $3d$ band shows $k$-independent hybridization with the VB, specifically, with the LH band.

Our picture of the IB formation and the associated ferromagnetism in (Ga,Mn)As may be naturally understood starting from the Anderson impurity model, which describes the transition-metal $d$ orbitals and the host band electrons hybridizing with each other. 
When hybridization between the $d$ orbital and the host ligand band is strong enough, this model predicts the formation of a {\it split-off Mn-impurity state} located above VBM for a Mn ion embedded in GaAs host \cite{PRB_01_Okabayashi}, where the Mn ion weakly bounds to a hole carrier. 
The schematic image is shown in Fig.~\ref{GMA_VB}(c). 
This picture is analogous to the Zhang-Rice singlet (ZRS) state in high-$T_{c}$ cuprates \cite{PRB_88_Zhang} and well explains the binding energy of the Mn acceptor level in GaAs:Mn (Mn concentration of $1.1 \times 10^{17}$ cm$^{-3}$), $\sim 110$ meV above VBM estimated from luminescence measurements \cite{PRB_74_Schairer, PRL_87_Schneider}. 
According to the percolation theory of BMP, the non-monotonic temperature dependence of the transport properties of (Ga,Mn)As is qualitatively explained \cite{PRB_03_Kaminski}, as due to the hole localization around the Mn ions. 
From cluster-model calculations of the Mn $2p$ core-level photoemission spectra \cite{PRB_98_Okabayashi}, the exchange interaction between the $d$ magnetic moment and carrier spin for the split-off state $N \beta$ has previously been estimated as $-1.2 \pm 0.2$ eV, where negative sign means antiferromagnetic magnetic interaction between the $3d$ moment and carrier spin. 
In this case, most of Mn $3d$ character hybridized into the VB goes to the split-off state and forms the IB, and the exchange splitting of the host GaAs band remains very weak, as observed by the resonant tunneling study \cite{NatPhys_11_Ohya, PRL_10_Ohya}. 
The local-density approximation calculations have predicted the existence of the IB as the antibonding states arising from the hybridization between Mn $3d$ state and ligand band \cite{APL_04_Mahadevan}, in contrast to the spin splitting of VB which is expressed as $\sim xN\beta \langle S \rangle$ in the picture of mean-field theory \cite{Science_00_Dietl, PRB_07_Jungwirth}. 
Following these arguments, we conclude that the formation of the IB is derived from the split-off state resulting in the ZRS-like BMP, with an overlap among the BMPs aligning the magnetic moments of the Mn ions in (Ga,Mn)As [Fig.~\ref{GMA_VB}(d)] and, therefore, the percolation of BMPs is the origin of the carrier-related ferromagnetism in (Ga,Mn)As. 
Generalizing our methodology, we have demonstrated that SX-ARPES allows observation of the Mn $3d$-derived IB in (Ga,Mn)As and the method will be also useful for other DMS systems. 
The promised physical information will deepen the fundamental understanding of the carrier-induced ferromagnetism in DMS which will help development of future spintronic device applications.

The authors thank C. Fadley, J. Min{\'{a}}r, and J. Kanski for informative discussions. This work was supported by a Grant-in-Aid for Scientific Research (S22224005 and 23000010) from JSPS, Japan. The authors thank C. Quitmann, F. van der Veen and J. Mesot for their continuous support of the SX-ARPES project at SLS. M.K. acknowledges support from the Japan Society for the Promotion of Science.

\end{document}